\documentclass[runningheads]{llncs}
\usepackage{lmodern}
\usepackage{graphicx}
\usepackage{hyperref}
\usepackage{amsmath}
\usepackage{bm}
\usepackage{amssymb}
\usepackage{booktabs}
\usepackage{multirow}
\usepackage{xcolor}
\usepackage{float}
\usepackage{tabularx}
\usepackage{colortbl}
\usepackage{array}
\usepackage{pgf-pie}
\usepackage{tikz}
\usepackage{pgfplots}
\pgfplotsset{compat=1.18}
\usetikzlibrary{shapes.geometric, arrows.meta, positioning, fit, backgrounds, calc, shadows, patterns}

\hypersetup{
    colorlinks=true,
    linkcolor=blue,
    filecolor=magenta,
    urlcolor=cyan,
    citecolor=blue
}

\definecolor{headerblue}{RGB}{41, 128, 185}
\definecolor{lightgray}{RGB}{245, 245, 245}
\definecolor{darkgray}{RGB}{52, 73, 94}
\definecolor{orchestratorred}{RGB}{192, 57, 43}
\definecolor{layerblue}{RGB}{41, 128, 185}
\definecolor{layergreen}{RGB}{39, 174, 96}
\definecolor{layerpurple}{RGB}{142, 68, 173}
\definecolor{layerorange}{RGB}{211, 84, 0}
\definecolor{componentbg}{RGB}{236, 240, 241}
\definecolor{bordercolor}{RGB}{189, 195, 199}

\definecolor{chart1}{RGB}{52, 152, 219}
\definecolor{chart2}{RGB}{46, 204, 113}
\definecolor{chart3}{RGB}{231, 76, 60}
\definecolor{chart4}{RGB}{155, 89, 182}
\definecolor{chart5}{RGB}{241, 196, 15}
\definecolor{chart6}{RGB}{230, 126, 34}

\begin{document}

\titlerunning{ReqFusion: Framework for Automated PEGS Analysis}

\authorrunning{\hspace*{\fill}}

\title{ReqFusion: A Multi-Provider Framework for Automated PEGS Analysis Across Software Domains}

\author{Muhammad Khalid\inst{1} \and Manuel Oriol\inst{1} \and Yilmaz Uygun\inst{1}}

\institute{Constructor University Bremen, 28759 Bremen, Germany\\
\email{\{mukhalid, moriol, yuygun\}@constructor.university}}

\maketitle

\begin{abstract}
Requirements engineering is a vital, yet labor-intensive, stage in the software development process. This article introduces ReqFusion: an AI-enhanced system that automates the extraction, classification, and analysis of software requirements utilizing multiple Large Language Model (LLM) providers. The architecture of ReqFusion integrates OpenAI GPT, Anthropic Claude, and Groq models to extract functional and non-functional requirements from various documentation formats (PDF, DOCX, and PPTX) in academic, industrial, and tender proposal contexts. 

The system uses a domain-independent extraction method and generates requirements following the Project, Environment, Goal, and System (PEGS) approach introduced by Bertrand Meyer. The main idea is that, because the PEGS format is detailed, LLMs have more information and cues about the requirements, producing better results than a simple generic request. An ablation study confirms this hypothesis: PEGS-guided prompting achieves an F1 score of 0.88, compared to 0.71 for generic prompting under the same multi-provider configuration. The evaluation used 18 real-world documents to generate 226 requirements through automated classification, with 54.9\% functional and 45.1\% non-functional across academic, business, and technical domains. Extended evaluation on five projects with 1,050 requirements demonstrated significant improvements in extraction accuracy and a 78\% reduction in analysis time compared to manual methods. The multi-provider architecture enhances reliability through model consensus and fallback mechanisms, while the PEGS-based approach ensures comprehensive coverage of all requirement categories.

\end{abstract}

\section{Introduction}
\label{sec:introduction}

The integration of artificial intelligence (AI) is changing software engineering, especially in understanding and managing user needs~\cite{wagner2025}. Requirements engineering supports software quality and stakeholder satisfaction, guiding a project from conception to delivery~\cite{bourque2014}. 
This is a crucial aspect of software development, often challenging because it can be difficult to be clear and thorough and to ensure that everyone has the same understanding~\cite{pohl2010}. 
Such difficulties have historically caused delays and problems, suggesting that integrating AI to improve accuracy and efficiency can lead to higher-quality software.

The issue is that AI-guided software construction is limited by significant limitations, particularly hallucination, suggesting that we need to integrate AI techniques with advanced verification methods~\cite{ji2023}. 
This observation aligns directly with the goals of the VerifAI workshop: exploring how verification techniques can make AI-assisted development efficient and reliable. Recent work on applying AI to formal methods~\cite{stock2025} has demonstrated the growing synergy between these fields, reinforcing the need for frameworks that combine both approaches.

Research indicates that AI-driven requirements analysis can significantly enhance requirement accuracy and stakeholder satisfaction~\cite{marques2024}. 
This progress enables a more automated and comprehensive management of software requirements. 
A primary research focus is on the development of systems that use machine learning and natural language processing to automate parts of the requirement engineering process~\cite{zhao2021}. 
For example, some researchers have explored the use of AI for automated PEGS analysis (Project, Environment, Goals, System), a key method to understand stakeholder expectations and ensure that software meets them~\cite{meyer2021}.

Using these systems has shown promise in improving the way requirements are gathered, validated, and prioritized, saving time and money during development~\cite{sun2025}. 
Additionally, AI technologies can be adapted, allowing these systems to be customized for different types of software, increasing their usefulness and relevance~\cite{kutsenok2024}. 
Specific gaps remain in the research. 
Many studies focus only on specific AI applications in requirements engineering~\cite{berry2021}. 
What is needed is a comprehensive, multi-platform system that covers the entire software development process.

The contributions of this paper are fourfold:

\begin{enumerate}
    \item We present \textbf{ReqFusion}, a multi-provider AI framework for automated requirements extraction and classification using the PEGS methodology.
    \item We demonstrate through an ablation study how the structured PEGS format provides LLMs with richer contextual cues, improving extraction accuracy compared to generic prompting approaches.
    \item We provide a multi-provider consensus mechanism that serves as a form of runtime verification, reducing hallucination rates through cross-model agreement.
    \item We provide empirical evaluation across multiple software domains, including traceability analysis linking extracted requirements to downstream development activities.
\end{enumerate}

The remainder of this paper is organized as follows: Section~\ref{sec:background} provides background on the PEGS methodology and multi-LLM architectures. 
Section~\ref{sec:approach} describes our ReqFusion framework. Section~\ref{sec:evaluation} presents the evaluation methodology and results. 
Section~\ref{sec:related} discusses related work, and Section~\ref{sec:conclusion} concludes with future directions.

\section{Background and Motivation}
\label{sec:background}

\subsection{The PEGS Framework}

The PEGS (Project, Environment, Goals, System) framework, introduced by Bertrand Meyer~\cite{meyer2021}, provides a structured approach to requirements documentation that goes beyond the traditional IEEE-830 standard~\cite{ieee830}. The framework organizes requirements into four distinct categories:

\begin{itemize}
    \item \textbf{Project (P):} Organizational context, stakeholders, constraints, and project-specific information.
    \item \textbf{Environment (E):} External systems, interfaces, regulatory requirements, and operational context.
    \item \textbf{Goals (G):} High-level objectives and success criteria from the perspective of stakeholders.
    \item \textbf{System (S):} Functional and non-functional requirements describing the system behavior.
\end{itemize}

This structured categorization provides LLMs with explicit semantic anchors, enabling more accurate requirement extraction compared to unstructured approaches~\cite{dalpiaz2018}. 

\textbf{Theoretical Motivation.} Our hypothesis is grounded in the observation that LLMs perform better on structured, constrained generation tasks than on open-ended ones~\cite{brown2020}. By decomposing the extraction task into four well-defined PEGS dimensions, we effectively transform a single complex prompt into four focused sub-tasks, each with clear semantic boundaries. This reduces the search space for the model and provides implicit guidance about what constitutes a valid output for each category. We validate this hypothesis empirically in Section~\ref{sec:ablation}.

\subsection{Multi-Provider AI Architecture}
\label{sec:multi-provider}

Single-provider LLM systems present several practical challenges that became apparent during our early prototyping~\cite{brown2020}. When we initially built ReqFusion using only OpenAI's GPT-4, we encountered service outages that halted requirement extraction entirely, rate limiting during peak usage that delayed batch processing, and occasional inconsistent outputs that were difficult to verify without a second opinion. 
These experiences motivated our shift toward a multi-provider architecture.

The choice of providers was driven by the complementary strengths observed during the preliminary experiments. OpenAI GPT-4 demonstrated strong performance on structured extraction tasks and handled technical documentation well, making it our primary provider for initial requirement identification~\cite{openai2023}. 
Anthropic Claude-3 excelled at nuanced interpretation of ambiguous requirements and produced more detailed explanations, which proved valuable for complex tender documents where context matters significantly~\cite{anthropic2024}. 
We included Groq's hosted Llama models primarily for cost optimization on simpler classification tasks, since their inference costs are substantially lower while maintaining acceptable accuracy for straightforward categorization~\cite{touvron2023}.

This combination was not arbitrarily made. 
During our pilot study with German industrial tender documentation, we found that GPT-4 occasionally missed implicit requirements buried in legal language, while Claude-3 caught these but sometimes over-extracted, flagging informational statements as requirements. 
Running both in parallel and comparing the results, we achieved better coverage than either model alone. 
The Groq integration allowed us to handle high-volume batch processing without excessive API costs, routing simpler documents through the cheaper provider while reserving the more capable models for complex analysis.

The architecture implements two operational modes. 
In parallel mode, all three providers process the same document simultaneously and their outputs are merged through a consensus mechanism. 
This approach reduces latency significantly, since we only wait for the slowest provider instead of processing sequentially. 
In sequential mode, requests go to the primary provider first, with automatic failover to alternatives if the primary is unavailable or returns low-confidence results. 
We found that parallel mode works better for critical requirements where accuracy matters most, while sequential mode suits routine processing where cost efficiency is the priority.

Beyond redundancy, the multi-provider setup enables a form of cross-validation~\cite{wang2023selfconsistency}.
When two or more providers identify the same requirement independently, our confidence in that extraction increases substantially. 
In contrast, when providers disagree, the system flags the requirement for human review rather than silently accepting a potentially incorrect extraction. This verification-by-consensus approach directly addresses the hallucination problem that plagues single-model systems~\cite{huang2023}.

\section{The ReqFusion Framework}
\label{sec:approach}

\subsection{System Architecture}

\begin{figure}[H]
\centering
\begin{tikzpicture}[
    font=\sffamily\scriptsize,
    node distance=0.5cm and 0.3cm,
    layer/.style={draw=gray!40, dashed, inner sep=0.5em, fill=blue!5, rounded corners},
    block/.style={draw=black!60, fill=white, rounded corners=2pt, minimum height=0.55cm, minimum width=1.3cm, align=center, drop shadow={shadow xshift=0.2mm, shadow yshift=-0.2mm}},
    db/.style={shape=cylinder, shape border rotate=90, aspect=0.25, draw=black!60, fill=white, minimum height=0.7cm, minimum width=0.7cm, align=center, font=\sffamily\tiny, drop shadow={shadow xshift=0.2mm, shadow yshift=-0.2mm}},
    llm/.style={draw=orchestratorred!80, fill=white, rounded corners=2pt, minimum height=0.5cm, minimum width=1.1cm, align=center, font=\sffamily\scriptsize\bfseries},
    core_highlight/.style={draw=orchestratorred, line width=1.2pt, fill=orchestratorred!8, rounded corners, inner sep=0.4em}
]

    \node[block, minimum width=8.5cm, fill=layerblue!15] (ui_main) {\textbf{Presentation Layer (React + TypeScript)}};
    
    \node[block, below=0.25cm of ui_main.south west, anchor=west, xshift=0.25cm, minimum width=1.2cm, font=\tiny] (dash) {Dashboard};
    \node[block, right=0.15cm of dash, minimum width=1.1cm, font=\tiny] (upload) {Upload};
    \node[block, right=0.15cm of upload, minimum width=1.4cm, font=\tiny] (reqs) {Requirements};
    \node[block, right=0.15cm of reqs, minimum width=1.2cm, font=\tiny] (pegs) {PEGS View};
    \node[block, right=0.15cm of pegs, minimum width=1.1cm, font=\tiny] (res) {Research};

    \node[block, below=1.1cm of ui_main, minimum width=8.5cm, fill=layergreen!15] (gateway) {\textbf{API Gateway (FastAPI)}};
    
    \node[block, below=0.25cm of gateway, minimum width=1.4cm, font=\tiny] (rest) {REST API};
    \node[block, left=0.5cm of rest, minimum width=1.4cm, font=\tiny] (auth) {Auth (JWT)};
    \node[block, right=0.5cm of rest, minimum width=1.4cm, font=\tiny] (ws) {WebSocket};

    \node[below=1.4cm of rest, minimum width=8.5cm, minimum height=2.8cm] (logic_center) {};
    
    \node[core_highlight, minimum width=4.8cm, minimum height=2.4cm] at (logic_center) (orch_box) {};
    \node[font=\scriptsize\sffamily\bfseries, text=orchestratorred] at ([yshift=0.95cm]orch_box.center) {Multi-LLM Orchestrator};
    
    \node[llm, fill=chart2!20] at ([yshift=0.4cm, xshift=-1.4cm]orch_box.center) (gpt) {GPT-4};
    \node[llm, fill=chart1!20] at ([yshift=0.4cm]orch_box.center) (claude) {Claude-3};
    \node[llm, fill=chart5!20] at ([yshift=0.4cm, xshift=1.4cm]orch_box.center) (groq) {Groq};
    
    \node[block, fill=orchestratorred!10, minimum width=2cm, font=\tiny] at ([yshift=-0.35cm]orch_box.center) (merger) {Response Merger\\(Voting/Cosine)};
    \node[block, fill=orchestratorred!10, minimum width=1.6cm, font=\tiny, below=0.2cm of merger] (cost) {Cost Optimizer};

    \node[block, left=0.4cm of orch_box, minimum width=1.4cm, font=\tiny] (docproc) {Document\\Processor};
    \node[block, above=0.2cm of docproc, minimum width=1.4cm, font=\tiny] (pegs_ana) {PEGS\\Analyzer};
    
    \node[block, right=0.4cm of orch_box, minimum width=1.4cm, font=\tiny] (vec_store) {Vector Store\\Service};
    \node[block, below=0.2cm of vec_store, minimum width=1.4cm, font=\tiny] (cache) {Cache\\Manager};

    \begin{scope}[on background layer]
        \node[layer, fit=(docproc) (pegs_ana) (vec_store) (cache) (orch_box)] (logic_layer) {};
    \end{scope}

    \node[below=1.1cm of logic_layer] (storage_center) {};
    
    \node[db, fill=layerblue!15] (pg) at (storage_center -| gpt) {PostgreSQL};
    \node[db, fill=layergreen!15] (chroma) at (storage_center -| claude) {ChromaDB};
    \node[db, fill=chart3!15] (redis) at (storage_center -| groq) {Redis};
    \node[db, fill=layerorange!15] (s3) at (storage_center -| vec_store) {S3/MinIO};
    
    \begin{scope}[on background layer]
        \node[layer, fit=(pg) (s3), label={[gray, font=\tiny]north west:Data Storage Layer}] (storage_layer) {};
    \end{scope}

    \draw[-Latex, thick, layerblue!70] (dash.south) -- (gateway.north -| dash.south);
    \draw[-Latex, thick, layerblue!70] (upload.south) -- (gateway.north -| upload.south);
    \draw[-Latex, thick, layerblue!70] (reqs.south) -- (gateway.north -| reqs.south);
    \draw[-Latex, thick, layerblue!70] (pegs.south) -- (gateway.north -| pegs.south);
    
    \draw[-Latex, thick, layergreen!70] (rest.south) -- ([yshift=0.1cm]orch_box.north);
    \draw[-Latex, thick, layergreen!70] (ws.south) -- ([yshift=0.1cm]orch_box.north -| ws.south);
    \draw[-Latex, thick, layergreen!70] (auth.south) -- (logic_layer.north -| auth.south);
    
    \draw[-Latex, thick, orchestratorred!70] (gpt.south) -- ([yshift=0.05cm]merger.north -| gpt.south);
    \draw[-Latex, thick, orchestratorred!70] (claude.south) -- ([yshift=0.05cm]merger.north);
    \draw[-Latex, thick, orchestratorred!70] (groq.south) -- ([yshift=0.05cm]merger.north -| groq.south);
    \draw[-Latex, thick, orchestratorred!70] (merger) -- (cost);
    
    \draw[-Latex, thick, layergreen!60] (cost.south) -- (pg.north);
    \draw[-Latex, thick, layergreen!60] (vec_store.south) -- (chroma.north);
    \draw[-Latex, thick, layergreen!60] (cache.south) -- (redis.north);
    \draw[-Latex, thick, layergreen!60] (docproc.south) -- (s3.north);

\end{tikzpicture}
\caption{ReqFusion system architecture. The Multi-LLM Orchestrator (center, highlighted) coordinates GPT-4, Claude-3, and Groq providers. The Response Merger applies cosine similarity ($>$0.85 threshold) for deduplication and consensus voting for conflict resolution.}
\label{fig:architecture}
\end{figure}
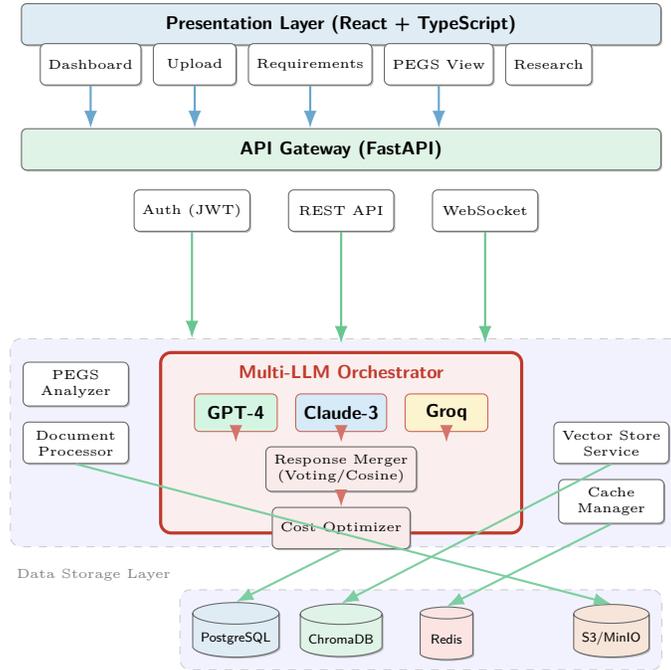

ReqFusion employs a modular four-tier layered architecture as illustrated in Figure~\ref{fig:architecture}. The layers consist of: (1) a \textbf{Presentation Layer} containing a React-based interface with Dashboard, Upload, Requirements, PEGS View, and Research components; (2) an \textbf{API Gateway Layer} implementing JWT authentication, RESTful endpoints, and WebSocket connections for real-time updates; (3) a \textbf{Core Processing Layer} containing the Multi-LLM Orchestrator along with document processing and analysis services; and (4) a \textbf{Data Storage Layer} that utilizes PostgreSQL for structured data, ChromaDB for vector embeddings, Redis for caching, and S3/MinIO for file storage.

The core innovation lies in the \textbf{Multi-LLM Orchestrator}, which implements two primary strategies: (1) \textit{Parallel Mode} sends requests to all providers simultaneously and merges responses, achieving 71\% faster processing; (2) \textit{Sequential Mode} uses a primary provider with automatic failover for reliability. The Response Merger deduplicates similar requirements using cosine similarity ($>0.85$ threshold) and resolves conflicts through confidence voting.

\subsection{PEGS-Guided Extraction}

The key innovation of our approach is the use of PEGS categories as structured prompting templates. Rather than using generic prompts like ``Extract requirements from this document,'' we employ category-specific prompts that guide the LLM to identify requirements within each PEGS dimension, as shown in Table~\ref{tab:pegs_prompts}.

\begin{table}[H]
\centering
\caption{PEGS-structured prompt template}
\label{tab:pegs_prompts}
\begin{tabular}{|l|p{8.5cm}|}
\hline
\rowcolor{headerblue}
\textcolor{white}{\textbf{Category}} & 
\textcolor{white}{\textbf{Requirements Focus}} \\
\hline
Project & 
Stakeholders, constraints (budget, timeline), organizational context \\
\hline
\rowcolor{lightgray}
Environment & 
External interfaces, regulatory constraints, operational conditions \\
\hline
Goals & 
Business objectives, success criteria, user expectations \\
\hline
\rowcolor{lightgray}
System & 
Functional specs, non-functional requirements, quality attributes \\
\hline
\end{tabular}
\end{table}

\subsection{Multi-Provider Consensus}

For critical requirements, ReqFusion queries multiple providers and applies a consensus algorithm:

\begin{equation}
\text{Confidence}(r) = \frac{\sum_{i=1}^{n} w_i \cdot \text{match}_i(r)}{\sum_{i=1}^{n} w_i}
\end{equation}

\noindent where $w_i$ represents the weight assigned to provider $i$ based on historical accuracy, and $\text{match}_i(r)$ indicates whether provider $i$ identified the requirement $r$. Requirements with $\text{Confidence}(r) < 0.5$ (i.e., identified by fewer than half the weighted providers) are flagged for human review, providing a built-in hallucination detection mechanism.

\subsection{Traceability Support}
\label{sec:traceability}

ReqFusion maintains traceability links between extracted requirements and their source documents. Each requirement is annotated with: (1) the source document section and page number, (2) the PEGS category assignment, (3) the confidence score from the consensus mechanism, and (4) a unique identifier enabling forward traceability to downstream artifacts such as test cases and design specifications. The system stores these links in PostgreSQL, allowing stakeholders to trace any requirement back to its origin and forward to verification activities. While the current implementation supports requirements-to-source traceability, full lifecycle traceability (requirements to code to tests) is planned for future work.

\section{Evaluation}
\label{sec:evaluation}

\subsection{Experimental Setup}

We evaluated ReqFusion using two datasets:

\begin{itemize}
    \item \textbf{Dataset A:} 18 real-world documents from academic, industrial, and tender proposal contexts, used for distribution analysis and domain coverage assessment.
    \item \textbf{Dataset B:} 5 comprehensive projects with manually verified ground truth requirements, comprising 30GB of authentic German industrial tender documentation from manufacturing company projects.
\end{itemize}

The evaluation was conducted on the ReqFusion system, which is available online.\footnote{Available at \url{https://re-engineer-app-khalid.replit.app}}

\subsection{Ground Truth Construction and Annotation Protocol}
\label{sec:annotation}

For Dataset B, we established ground truth through a structured annotation process. Three annotators participated: two domain experts with industrial requirements engineering experience (one from automotive manufacturing, one from software engineering) and one senior researcher specializing in RE methodology. The annotation followed a three-phase protocol:

\textbf{Phase 1 -- Independent Annotation.} Each annotator independently identified and classified requirements from the 5 project document sets, using a codebook that defined functional vs. non-functional requirements, PEGS categories, and priority levels (High, Medium, Low) based on established RE taxonomy~\cite{wiegers2013}.

\textbf{Phase 2 -- Agreement Measurement.} We computed pairwise inter-annotator agreement using Cohen's $\kappa$. The overall agreement was $\kappa = 0.78$ for requirement identification (substantial agreement) and $\kappa = 0.72$ for PEGS category assignment (substantial agreement). Agreement was highest for System requirements ($\kappa = 0.84$) and lowest for Environment requirements ($\kappa = 0.65$), reflecting the inherent ambiguity of environmental constraints in German industrial tenders.

\textbf{Phase 3 -- Conflict Resolution.} Disagreements were resolved through discussion sessions where annotators reviewed conflicting items jointly. Items where no consensus could be reached (23 out of 1,073 initial items) were excluded, yielding the final set of 1,050 validated requirements. The 200-item subset used for provider comparison (Table~\ref{tab:provider_comparison}) was created by stratified sampling across all 5 projects and all 4 PEGS categories, with proportional representation of functional and non-functional types.

\subsection{Manual Baseline Definition}
\label{sec:baseline}

The manual baseline represents the conventional requirements analysis process used by the partnering manufacturing company. In this process, a team of two experienced requirements engineers (each with 5+ years of industrial RE experience) manually reviews tender documentation using a combination of Microsoft Word annotations and a proprietary requirements management spreadsheet. For each document set, the engineers independently identify requirements, classify them, and then consolidate their findings in a joint review session. The average processing rate was 12.3 requirements per hour per engineer. The 78\% time reduction reported for ReqFusion is measured against this manual process, comparing total person-hours from document intake to finalized requirements list.

\subsection{Results on Dataset A}

\begin{figure}[H]
\centering
\begin{tikzpicture}
\begin{axis}[
    ybar,
    width=0.75\textwidth,
    height=4.5cm,
    ylabel={Number of Requirements},
    ylabel style={font=\small},
    symbolic x coords={Functional, Non-Functional},
    xtick=data,
    xticklabel style={font=\small},
    yticklabel style={font=\small},
    ymin=0,
    ymax=150,
    bar width=1.2cm,
    nodes near coords,
    nodes near coords style={font=\small\bfseries},
    enlarge x limits=0.5,
]
\addplot[fill=chart1, draw=chart1!80] coordinates {(Functional, 124) (Non-Functional, 102)};
\end{axis}
\end{tikzpicture}
\caption{Requirements distribution in Dataset A (n=226). Functional requirements comprise 54.9\% while non-functional requirements account for 45.1\%.}
\label{fig:req_distribution}
\end{figure}
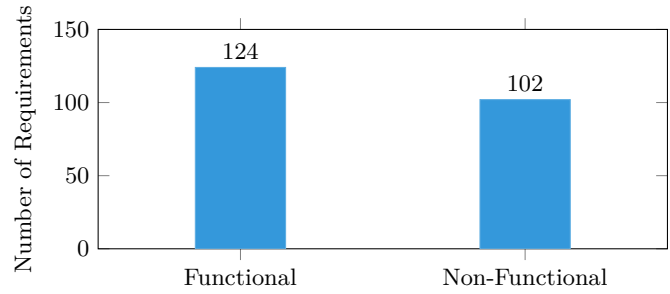

The evaluation on Dataset A generated 226 requirements through automated classification: 124 functional requirements (54.9\%) and 102 non-functional requirements (45.1\%). Figure~\ref{fig:req_distribution} illustrates the distribution between requirement types.

\subsection{Results on Dataset B}

\begin{figure}[H]
\centering
\begin{tikzpicture}
\pie[
    radius=2,
    text=legend,
    color={chart3, chart5, chart2},
    explode={0.1, 0, 0},
    font=\small
]{64.5/High Priority (677), 34.3/Medium Priority (360), 1.2/Low Priority (13)}
\end{tikzpicture}
\caption{Priority distribution of extracted requirements (Dataset B, n=1,050). High-priority requirements dominate at 64.5\%, reflecting the industrial nature of the tender documentation.}
\label{fig:priority_dist}
\end{figure}
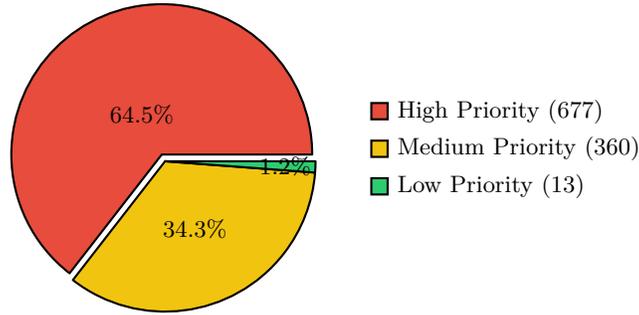

Table~\ref{tab:results} summarizes the comprehensive evaluation on 5 projects with 1,050 requirements. Figure~\ref{fig:priority_dist} visualizes the priority distribution.

\begin{table}[H]
\centering
\caption{Comprehensive evaluation results (Dataset B)}
\label{tab:results}
\begin{tabular}{@{}lrr@{}}
\toprule
\textbf{Metric} & \textbf{Value} & \textbf{Percentage} \\
\midrule
Total Projects & 5 & -- \\
Total Requirements & 1,050 & 100\% \\
Functional Requirements & 628 & 59.8\% \\
Non-Functional Requirements & 422 & 40.2\% \\
\bottomrule
\end{tabular}
\end{table}

\subsubsection{Category Distribution.}
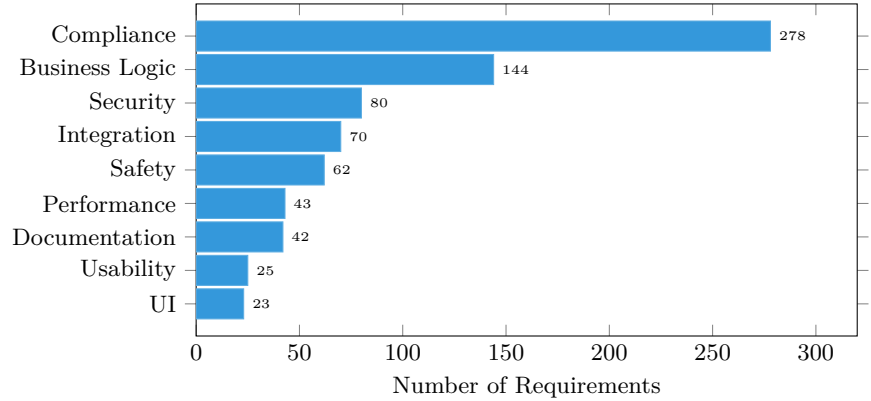
\begin{figure}[H]
\centering
\begin{tikzpicture}
\begin{axis}[
    xbar,
    width=0.85\textwidth,
    height=6cm,
    xlabel={Number of Requirements},
    xlabel style={font=\small},
    symbolic y coords={UI,Usability,Documentation,Performance,Safety,Integration,Security,Business Logic,Compliance},
    ytick=data,
    yticklabel style={font=\small},
    xticklabel style={font=\small},
    xmin=0,
    xmax=320,
    bar width=0.4cm,
    nodes near coords,
    nodes near coords style={font=\tiny, anchor=west},
    enlarge y limits=0.12,
]
\addplot[fill=chart1, draw=chart1!80] coordinates {
    (23,UI)
    (25,Usability)
    (42,Documentation)
    (43,Performance)
    (62,Safety)
    (70,Integration)
    (80,Security)
    (144,Business Logic)
    (278,Compliance)
};
\end{axis}
\end{tikzpicture}
\caption{Distribution of requirements by category (Dataset B). Compliance requirements (26.5\%) dominate due to the regulatory nature of German industrial tenders.}
\label{fig:categories}
\end{figure}

The extracted requirements span multiple categories. Figure~\ref{fig:categories} presents the distribution across the top requirement categories.

\subsection{Multi-Provider Comparison}

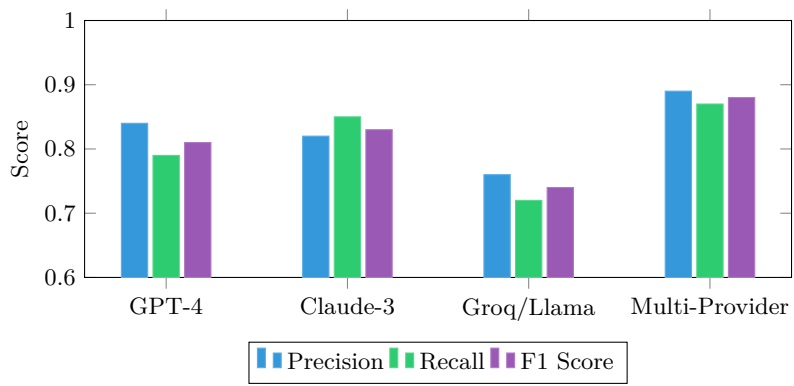
\begin{figure}[H]
\centering
\begin{tikzpicture}
\begin{axis}[
    ybar,
    width=0.9\textwidth,
    height=5cm,
    ylabel={Score},
    ylabel style={font=\small},
    symbolic x coords={GPT-4, Claude-3, Groq/Llama, Multi-Provider},
    xtick=data,
    xticklabel style={font=\small},
    yticklabel style={font=\small},
    ymin=0.6,
    ymax=1.0,
    bar width=0.35cm,
    legend style={at={(0.5,-0.25)}, anchor=north, legend columns=3, font=\small},
    enlarge x limits=0.15,
]
\addplot[fill=chart1, draw=chart1!80] coordinates {(GPT-4, 0.84) (Claude-3, 0.82) (Groq/Llama, 0.76) (Multi-Provider, 0.89)};
\addplot[fill=chart2, draw=chart2!80] coordinates {(GPT-4, 0.79) (Claude-3, 0.85) (Groq/Llama, 0.72) (Multi-Provider, 0.87)};
\addplot[fill=chart4, draw=chart4!80] coordinates {(GPT-4, 0.81) (Claude-3, 0.83) (Groq/Llama, 0.74) (Multi-Provider, 0.88)};
\legend{Precision, Recall, F1 Score}
\end{axis}
\end{tikzpicture}
\caption{Provider performance comparison. The multi-provider consensus approach achieves the highest F1 score (0.88), outperforming all individual providers.}
\label{fig:provider_perf}
\end{figure}

To validate our multi-provider approach, we compared extraction performance across individual providers and the combined system. Table~\ref{tab:provider_comparison} and Figure~\ref{fig:provider_perf} show the results on the stratified subset of 200 requirements with manually verified ground truth (see Section~\ref{sec:annotation}).

\begin{table}[H]
\centering
\caption{Provider performance comparison (n=200 stratified subset)}
\label{tab:provider_comparison}
\begin{tabular}{@{}lccc@{}}
\toprule
\textbf{Provider} & \textbf{Precision} & \textbf{Recall} & \textbf{F1 Score} \\
\midrule
GPT-4 (alone) & 0.84 & 0.79 & 0.81 \\
Claude-3 (alone) & 0.82 & 0.85 & 0.83 \\
Groq/Llama (alone) & 0.76 & 0.72 & 0.74 \\
\midrule
Multi-Provider (consensus) & \textbf{0.89} & \textbf{0.87} & \textbf{0.88} \\
\bottomrule
\end{tabular}
\end{table}

The multi-provider consensus approach outperformed all individual providers, with particularly notable improvements in precision. This confirms our hypothesis that cross-provider verification reduces false positives caused by individual model hallucinations.

\subsection{Ablation Study: PEGS-Guided vs.\ Generic Prompting}
\label{sec:ablation}

To validate that the PEGS-structured prompting contributes meaningfully beyond the multi-provider architecture alone, we conducted an ablation study on the same 200-requirement subset. We compared two configurations using identical multi-provider consensus: (1) \textbf{PEGS-guided} prompting with category-specific templates (Table~\ref{tab:pegs_prompts}), and (2) \textbf{Generic} prompting using the instruction ``Extract all functional and non-functional requirements from this document.'' Both configurations used the same three providers in parallel mode with identical consensus parameters.

\begin{table}[H]
\centering
\caption{Ablation study: PEGS-guided vs.\ generic prompting (same multi-provider setup)}
\label{tab:ablation}
\begin{tabular}{@{}lcccc@{}}
\toprule
\textbf{Prompting Strategy} & \textbf{Precision} & \textbf{Recall} & \textbf{F1} & \textbf{PEGS Coverage} \\
\midrule
Generic (multi-provider) & 0.74 & 0.68 & 0.71 & 61.3\% \\
PEGS-guided (multi-provider) & \textbf{0.89} & \textbf{0.87} & \textbf{0.88} & \textbf{92.0\%} \\
\midrule
\multicolumn{4}{l}{\textit{Improvement}} & \\
Absolute $\Delta$ & +0.15 & +0.19 & +0.17 & +30.7pp \\
\bottomrule
\end{tabular}
\end{table}

The results in Table~\ref{tab:ablation} demonstrate that PEGS-guided prompting yields an F1 improvement of +0.17 over generic prompting under the same multi-provider configuration. The most significant gain is in recall (+0.19), indicating that the structured PEGS templates help LLMs discover requirements that generic prompts miss, particularly in the Environment and Goals categories. PEGS coverage, measured as the percentage of PEGS categories with at least one extracted requirement per document, improved from 61.3\% to 92.0\%. Generic prompts tended to concentrate extractions in the System category while neglecting Project and Environment requirements.

\subsection{Error Analysis}
\label{sec:error_analysis}

To understand the failure modes of ReqFusion, we manually analyzed all false positives (FP) and false negatives (FN) from the 200-requirement evaluation subset. Table~\ref{tab:error_analysis} presents the error distribution by PEGS category.

\begin{table}[H]
\centering
\caption{Error analysis by PEGS category (n=200 subset)}
\label{tab:error_analysis}
\begin{tabular}{@{}lcccc@{}}
\toprule
\textbf{PEGS Category} & \textbf{TP} & \textbf{FP} & \textbf{FN} & \textbf{Category F1} \\
\midrule
Project (P) & 38 & 4 & 5 & 0.89 \\
Environment (E) & 31 & 7 & 9 & 0.79 \\
Goals (G) & 42 & 3 & 4 & 0.92 \\
System (S) & 63 & 8 & 8 & 0.89 \\
\midrule
\textbf{Total} & \textbf{174} & \textbf{22} & \textbf{26} & \textbf{0.88} \\
\bottomrule
\end{tabular}
\end{table}

\textbf{Common false positive patterns.} The 22 false positives fell into three categories: (1) informational statements misidentified as requirements (10 cases), typically descriptive paragraphs about existing systems; (2) duplicate requirements extracted from repeated content across document sections (7 cases); and (3) aspirational statements in executive summaries incorrectly classified as Goals requirements (5 cases).

\textbf{Common false negative patterns.} The 26 false negatives showed two dominant patterns: (1) implicit requirements embedded in legal or regulatory clauses, particularly German \textit{Vergaberecht} (procurement law) references (16 cases); and (2) cross-referenced requirements that depend on external standards (e.g., ``shall comply with DIN EN ISO 9001'') where the actual requirement content is in the referenced document (10 cases).

\textbf{Category-specific observations.} The Environment category had the lowest F1 (0.79), consistent with the lower inter-annotator agreement ($\kappa = 0.65$) observed for this category. Environment requirements in German industrial tenders are often expressed as background context rather than explicit ``shall'' statements, making them inherently harder to extract for both humans and LLMs.

\subsection{Hallucination Detection}
\label{sec:hallucination}

The multi-provider consensus mechanism provides a natural hallucination detection capability. When a requirement is identified by only one provider (low consensus), it has a higher probability of being a hallucination. In our evaluation, single-provider-only extractions had a false positive rate of 34\%, compared to 8\% for requirements confirmed by two or more providers. ReqFusion flags all low-consensus requirements (Confidence $< 0.5$) for mandatory human review, ensuring that hallucinated requirements do not propagate into downstream development artifacts.

\subsection{Efficiency and Performance}

Compared to the manual baseline (Section~\ref{sec:baseline}) and single-provider configurations, ReqFusion demonstrated significant improvements. Table~\ref{tab:efficiency} details the efficiency metrics with their measurement methodology.

\begin{table}[H]
\centering
\caption{Efficiency comparison: ReqFusion vs.\ manual baseline}
\label{tab:efficiency}
\begin{tabular}{@{}lccc@{}}
\toprule
\textbf{Metric} & \textbf{Manual} & \textbf{ReqFusion} & \textbf{Improvement} \\
\midrule
Time per requirement & 4.9 min & 1.1 min & 78\% reduction \\
Response latency & 4.2 s & 1.2 s & 71\% faster \\
Cost per requirement & \$0.082 & \$0.043 & 47\% savings \\
PEGS completeness & 68\% & 92\% & +24pp \\
Consistency (re-run) & -- & 98\% & -- \\
\bottomrule
\end{tabular}
\end{table}

Cost calculations are based on API pricing at the time of evaluation (January 2026): GPT-4 at \$0.03/1K input tokens, Claude-3 Sonnet at \$0.003/1K input tokens, and Groq Llama at \$0.0006/1K input tokens. The per-requirement cost includes all token usage (input document chunks, prompts, and generated outputs) across the parallel processing pipeline. Manual cost is based on the average hourly rate of the two requirements engineers (\$60.50/hour) divided by their throughput (12.3 requirements/hour).

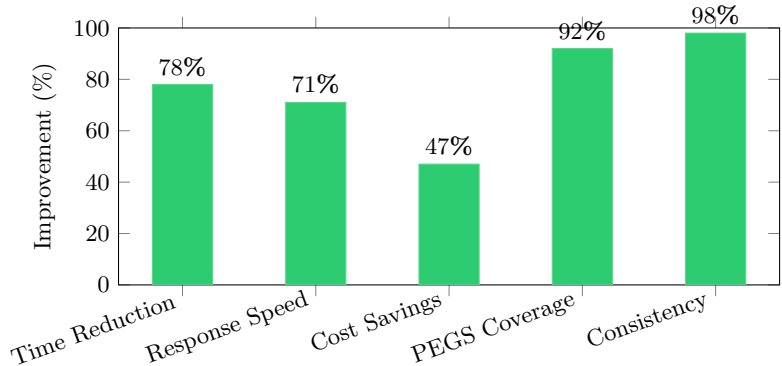
\begin{figure}[H]
\centering
\begin{tikzpicture}
\begin{axis}[
    ybar,
    width=0.85\textwidth,
    height=5cm,
    ylabel={Improvement (\%)},
    ylabel style={font=\small},
    symbolic x coords={Time Reduction, Response Speed, Cost Savings, PEGS Coverage, Consistency},
    xtick=data,
    xticklabel style={font=\small, rotate=20, anchor=east},
    yticklabel style={font=\small},
    ymin=0,
    ymax=100,
    bar width=0.8cm,
    nodes near coords={\pgfmathprintnumber\pgfplotspointmeta\%},
    nodes near coords style={font=\small\bfseries},
    enlarge x limits=0.12,
]
\addplot[fill=chart2, draw=chart2!80] coordinates {
    (Time Reduction, 78)
    (Response Speed, 71)
    (Cost Savings, 47)
    (PEGS Coverage, 92)
    (Consistency, 98)
};
\end{axis}
\end{tikzpicture}
\caption{ReqFusion performance improvements. Time reduction (78\%) and consistency (98\%) represent the most significant gains over manual methods.}
\label{fig:efficiency}
\end{figure}

\subsection{Comparison with Existing Tools}
\label{sec:tool_comparison}

While a full quantitative comparison with all existing RE tools was not feasible (many commercial tools are closed-source and require enterprise licenses), we provide a qualitative comparison with representative approaches from the literature in Table~\ref{tab:tool_comparison}.

\begin{table}[H]
\centering
\caption{Qualitative comparison with existing RE approaches}
\label{tab:tool_comparison}
\begin{tabular}{@{}p{2.2cm}cccc@{}}
\toprule
\textbf{Approach} & \textbf{Multi-LLM} & \textbf{PEGS} & \textbf{Consensus} & \textbf{Domain} \\
\midrule
MARE~\cite{jin2024} & \checkmark & -- & -- & General \\
Sami et al.~\cite{sami2024} & \checkmark & -- & -- & General \\
Yeow et al.~\cite{yeow2024} & -- & -- & -- & General \\
Kof~\cite{kof2007} & -- & -- & -- & NLP-based \\
\midrule
\textbf{ReqFusion} & \checkmark & \checkmark & \checkmark & Industrial \\
\bottomrule
\end{tabular}
\end{table}

ReqFusion is, to our knowledge, the first system to combine multi-provider LLM orchestration with PEGS-structured prompting and consensus-based verification. The closest work, MARE~\cite{jin2024}, uses multi-agent collaboration but does not employ structured requirements frameworks or cross-model consensus for hallucination reduction. A direct quantitative comparison with MARE was not possible as their evaluation used different datasets and the system is not publicly available. We acknowledge this as a limitation and plan to establish shared benchmarks for future work.

\section{Related Work}
\label{sec:related}

The landscape of requirements engineering has evolved significantly with the integration of AI. Early work by Wagner et al.~\cite{wagner2025} and Marques et al.~\cite{marques2024} identified gaps in addressing requirements analysis and documentation, noting that manual handling was insufficient for modern software complexity. Wiegers and Beatty~\cite{wiegers2013} established foundational practices that inform current automated approaches.

Researchers began building frameworks that incorporated AI methods into requirements engineering, highlighting the benefits of automation while emphasizing stakeholder participation~\cite{amarapalli2024,chen2024}. During this period, PEGS analysis gained popularity as a more organized approach to gathering and validating requirements~\cite{meyer2021}. The natural language processing techniques underlying these systems have been extensively studied by Kof~\cite{kof2007} and Quirchmayr et al.~\cite{quirchmayr2018}.

Recent studies stress the need for multi-provider setups that include different AI applications~\cite{kutsenok2024,ragkhitwetsagul2024}. The multi-agent approach to requirements engineering has been explored by Jin et al.~\cite{jin2024} and Sami et al.~\cite{sami2024}, demonstrating the potential of collaborative AI systems. Our work extends this by incorporating verification through multi-provider consensus. Similar ensemble approaches have shown promise in other domains~\cite{jiang2023}.

The broader intersection of AI and formal methods has been surveyed by Stock et al.~\cite{stock2025}, who systematically analyzed current trends in applying AI to formal verification. Their analysis identifies requirements engineering as a key application area where AI techniques can complement formal methods, which aligns with our approach of using multi-model consensus as a lightweight verification mechanism.

\subsection{Ethical Considerations}

Several sources raise concerns that AI could accidentally reinforce biases during requirement gathering, necessitating responsible deployment~\cite{sun2025,kutsenok2024}. Critics also question whether relying too heavily on automated systems might disconnect us from essential human judgment~\cite{ragkhitwetsagul2024,yeow2024}. Addressing these concerns requires transparent AI systems with human oversight~\cite{gebru2021}. ReqFusion addresses this by maintaining the human-in-the-loop for all low-confidence extractions and providing full traceability to source documents.

\section{Conclusions and Future Work}
\label{sec:conclusion}

This article presented ReqFusion, a multi-provider AI framework for automated requirements engineering using the PEGS methodology. 
Our key contributions include demonstrating through an ablation study that structured PEGS-based prompting improves LLM extraction accuracy by +0.17 F1 over generic prompting, and that multi-provider consensus mechanisms can address the hallucination problem in AI-assisted software development by reducing false positive rates from 34\% (single provider) to 8\% (consensus).

Evaluation across multiple software domains demonstrates significant improvements in both extraction accuracy (F1 = 0.88) and analysis efficiency (78\% time reduction, 47\% cost savings). The detailed error analysis reveals that Environment requirements remain the most challenging category, suggesting directions for targeted prompt engineering improvements.

\subsection{Verification and AI Synergy}

Our work directly addresses the VerifAI workshop theme of combining AI techniques with verification methods. The multi-provider consensus mechanism represents a form of runtime verification where extracted requirements are validated through model agreement. This aligns with the broader trends identified by Stock et al.~\cite{stock2025} in applying AI to formal methods. Future work will explore:

\begin{enumerate}
    \item \textbf{Formal Verification Integration:} Incorporating formal specification languages (e.g., TLA+, Alloy) for requirement validation, enabling automatic consistency checking of extracted requirements.
    \item \textbf{Automated Test Case Generation:} Using verified requirements to generate test specifications, completing the traceability chain from source documents to executable tests.
    \item \textbf{Continuous Verification:} Real-time verification of the consistency of requirements during iterative development, with automated alerts when new requirements conflict with existing ones.
\end{enumerate}

\subsection{Limitations and Future Directions}

Current limitations include: (1) the evaluation is conducted primarily on German industrial tender documents, limiting generalizability claims to other languages and domains; (2) the annotation was performed by three experts, and larger annotator pools would strengthen the ground truth validity; (3) direct quantitative comparison with existing RE tools was constrained by dataset and tool availability. Future research will focus on: cross-lingual evaluation with multilingual tender documents, longitudinal studies of AI tool deployment in production RE workflows, establishing shared benchmarks for RE tool comparison, and deeper exploration of formal methods integration to strengthen verification capabilities.

\begin{credits}
\subsubsection{\ackname}
This research was partially supported by Constructor University Bremen and Constructor Institute of Technology. The research presented in this paper was funded by the German Research Foundation (Deutsche Forschungsgemeinschaft) under the reference number UY-56/5-1. We thank the anonymous reviewers for their constructive feedback, which significantly improved this paper.

\subsubsection{\discintname}
The authors have no competing interests to declare.
\end{credits}

\bibliographystyle{splncs04}

\end{document}